\documentclass[11pt]{article}
\usepackage{moriond,epsfig,amsmath,amssymb}

\def\Journal#1#2#3#4{{#1} {\bf #2}, #3 (#4)}

\def\PRL{\em Phys. Rev. Lett.}

\def\ZP{\em Z. Phys.}
\def\PL{\em Phys. Lett.}
\def\PR{\em Phys. Rev.}
\def\EPJ{\em Eur. Phys. J.}

\def\be{\begin{equation}}
\def\ee{\end{equation}}
\def\bea{\begin{eqnarray}}
\def\eea{\end{eqnarray}}

\begin{document}
\vspace*{4cm}
\title{Bose-Einstein correlations in WW pair production at LEP}

\author{ N. van Remortel }

\address{Department of Physics, Universiteit Antwerpen\\ Universiteitsplein 1,
B-2610 Antwerpen, Belgium\\ email: Nick.van.Remortel@cern.ch}

\maketitle\abstracts{
The search for the presence of Bose-Einstein correlations (BEC) between
identical bosons coming from the decay of different $W$'s in the reaction
$e^+e^-\rightarrow W^+W^- \rightarrow q_1\bar{q_2}q_3\bar{q_4}$ is recently
of high interest. It is relevant for a precise determination of the $W$ mass
at LEP in the fully-hadronic decay channel and can provide insight in 
hadronisation mechanisms. A novel method, allowing data to be directly
compared with itself by means of mixing tracks from different events, is
applied by all LEP experiments. The latest results obtained by the L3 and
DELPHI collaboration are presented.}

\section{Introduction}
Correlations between final state particles in high energy collisions have
been extensively studied since the Goldhaber et al. experiment~\cite{goldhaber}.
They can be due to phase space, energy-momentum conservation, resonance
production, or are dynamical in nature. In the particular case of identical
bosons, the correlations are enhanced by the Bose-Einstein effect. These
Bose-Einstein Correlations (BEC) are a consequence of quantum statistics. The
net result is that multiplets of identical bosons are produced with smaller
energy-momentum differences than non-identical ones.

The interest of studying BEC in the reaction $e^+e^- \rightarrow W^+W^-
\rightarrow q_1\bar{q_2}q_3\bar{q_4}$, and more in particular BEC between
identical bosons coming from different $W$'s is twofold.
Due to the short life-time of a $W$ boson ($\tau c \sim$ 0.1 fm), compared to
the distance needed to produce final state particles (0.5 - 1.~fm), it is
natural to expect a large space-time overlap between the decay products of
the $W$'s. As a consequence, BEC could exist between identical bosons coming
from different $W$'s, hence violating the assumption that the two $W$'s
decay independently.  Together with colour reconnection~\cite{cr1}, the
poor understanding of the inter-$W$ BEC effect introduces a large systematic
uncertainty in the measurement of the $W$ mass by means of direct
reconstruction in the fully-hadronic decay channel~\cite{bec1}. The current
statistical uncertainty of the combined LEP measurement in this channel
amounts to 36 MeV~\cite{lepewpaper}, while the systematic uncertainty due to
the inter-$W$ BEC effect is of the same order.
In addition, the observation of a possible inter-$W$ BEC effect is of high
interest to our understanding of hadronisation models, since, at present, all
assume independent hadronisation of the $W$ decay products.

After five successful years of LEP2 running, each of the four LEP experiments
collected around 10000 $WW$ events at centre-of-mass energies ranging from
189 to 209 GeV, nevertheless a measurement of inter-$W$ BEC
is mainly bound by statistics. The most promising technique, based
on~\cite{eddi1} and~\cite{eddi2} is currently applied by all four LEP
experiments. The L3 collaboration was the first to publish their results
obtained by this method~\cite{l3paper}. DELPHI has nearly finalised their
analysis while activities in ALEPH and OPAL are still ongoing.
Therefore, in what follows, only the results obtained by L3 and DELPHI will
be discussed.

\section{Formalism}
The method proposed in~\cite{eddi1,eddi2} provides an elegant way to
investigate the presence of inter-$W$ BEC. It allows real data to be directly
compared with itself, hereby eliminating fragmentation and detector
uncertainties, and is based on the following reasoning.

In the case of two stochastically independent hadronically decaying $W$'s,
the single and two-particle inclusive densities obey the following relations:
\begin{align}
\rho^{WW}(1) &= \rho^{W^+}(1)+\rho^{W^-}(1), \\
\rho^{WW}(1,2) &= 2\rho^{W}(1,2)+2\rho^{W^+}(1)\rho^{W^-}(2),
\label{densities}
\end{align}  
\noindent
where $\rho^{W}(1)$ denotes the inclusive single particle density of one $W$ and $\rho^{W}(1,2)$
the inclusive two-particle density of one $W$. The densities $\rho^{WW}(1)$ and
$\rho^{WW}(1,2)$ then correspond to the single, resp. two-particle inclusive
density of a fully hadronic $WW$ event.
The terms $\rho^{WW}(1,2)$ and $\rho^{W}(1,2)$ can be extracted from,
respectively, fully-hadronic and semi-leptonic $WW$ decays. The product of the
single particle densities $\rho^{W}(1)\rho^{W}(2)$ is, in practical
applications, replaced by a
two-particle density $\rho^{WW}_{\rm mix}$, obtained by combining particles
from two hadronic $W$ decays taken from different semi-leptonic events. 

Expressed in the variable $Q=\sqrt{-(p_1-p_2)^2}$, Eq.~\ref{densities} can be
re-written as
\begin{equation}
\rho^{WW}(Q) = 2\rho^{W}(Q)+2\rho^{WW}_{\rm mix}(Q).
\label{msimpler}
\end{equation}
Using this expression one can construct two test observables to search for
correlations between decay products from different $W$'s. The observables
considered are:
\begin{align}
\Delta \rho (Q) &= \rho^{WW}(Q) - 2\rho^{W}(Q) - 2\rho^{WW}_{mix}(Q), \label{drho}\\
D(Q) &= \frac{\rho^{WW}(Q)}{2\rho^{W}(Q) + 2\rho^{WW}_{mix}(Q)}. \label{d}
\end{align}
\noindent
Any deviation from zero of $\Delta \rho(Q)$, or any deviation from one of
$D(Q)$, is an indication for non-independent $WW$ decay. 
\section{L3 results}
\begin{figure}
\begin{center}
  \begin{tabular}{cc}
      \mbox{\psfig{figure=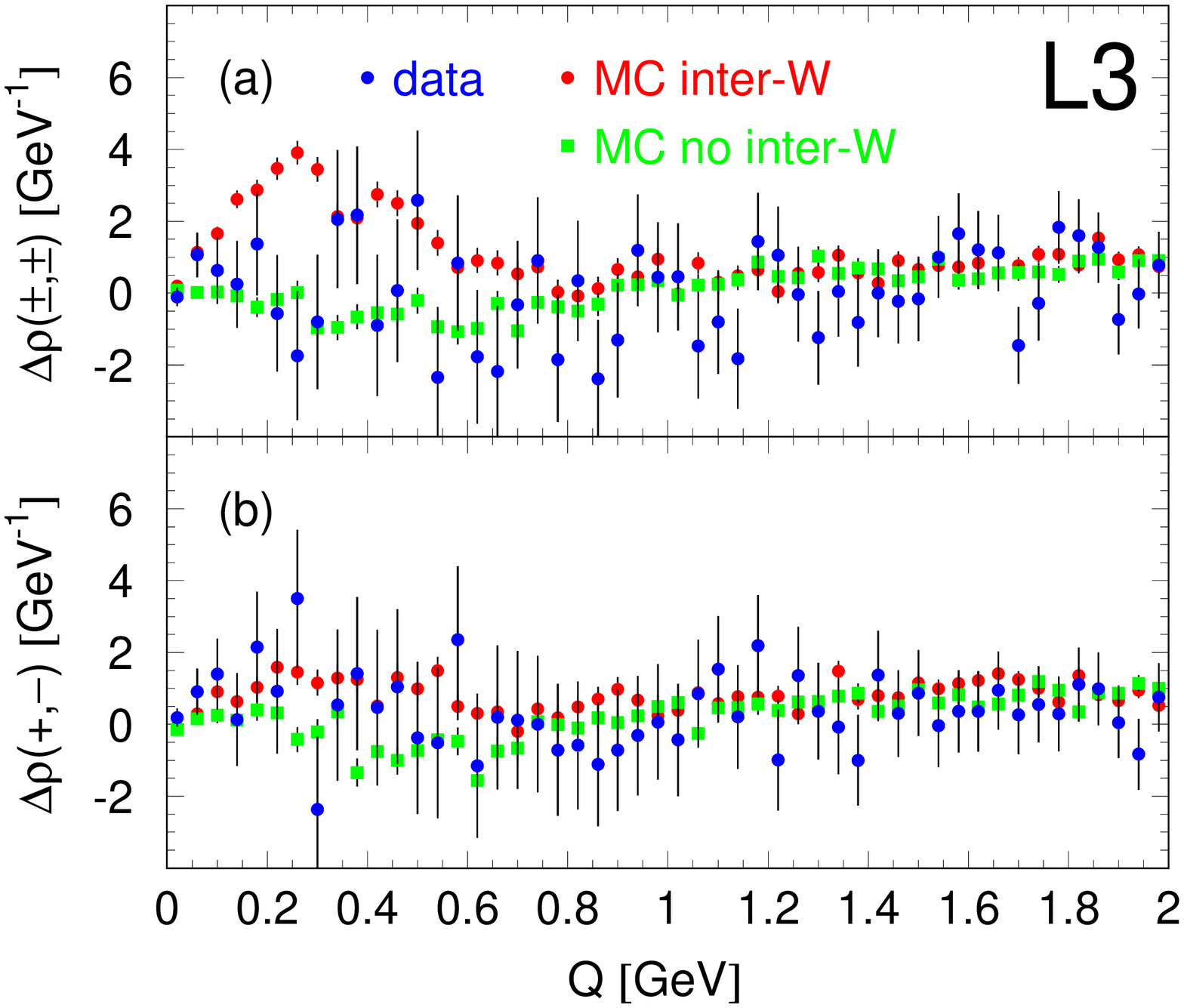,width=7.0cm}}
&
       \mbox{\psfig{figure=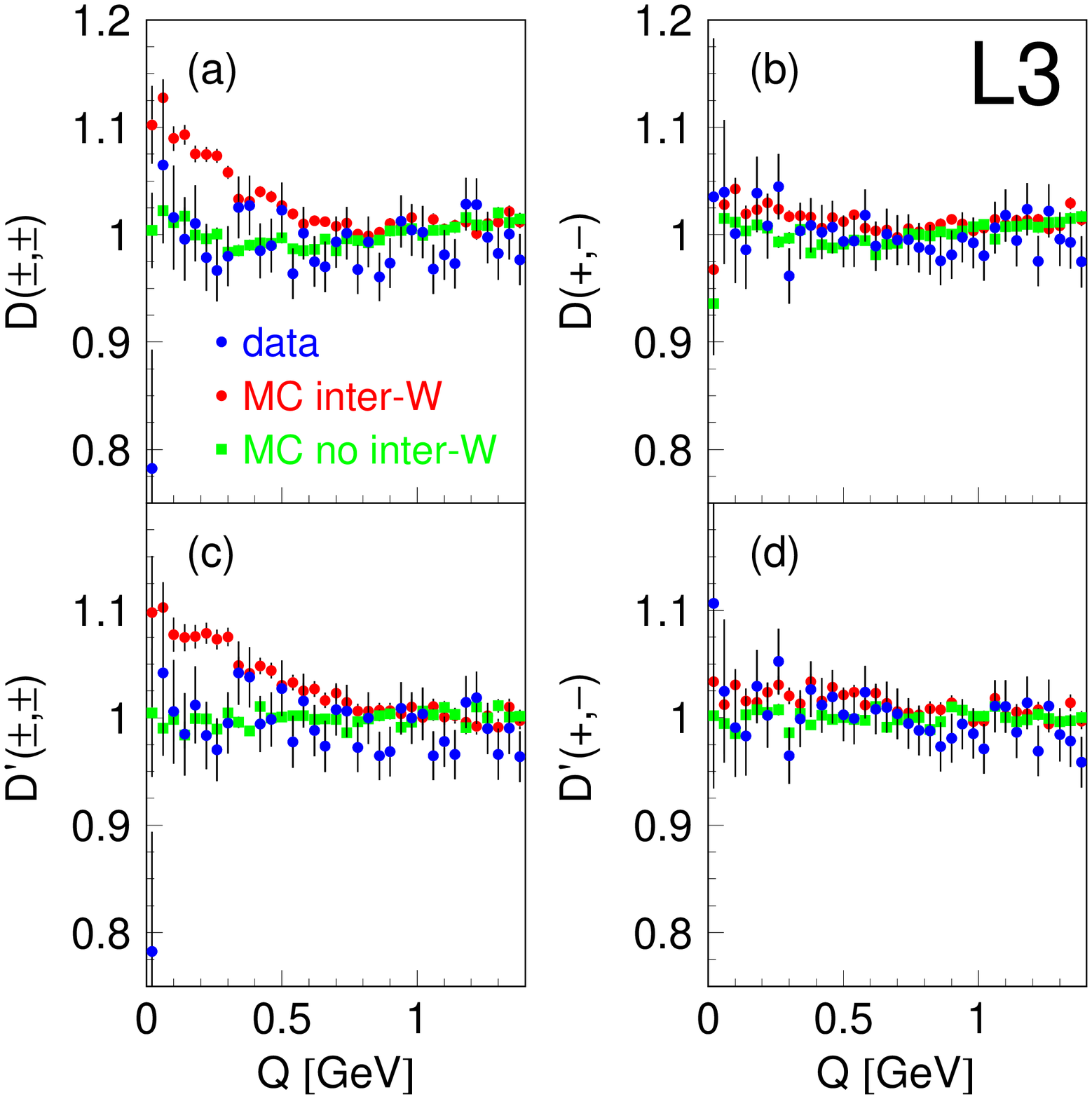,width=7.0cm}}
  \end{tabular}
\end{center}
\caption{The $\Delta \rho(Q)$ distribution (left) and the $D(Q)$ and $D'(Q)$
  distributions (right) for like-sign and unlike-sign particle pairs,
  obtained by the L3 experiment.
\label{fig:l3fig}}
\end{figure}

The L3 collaboration has recently published results~\cite{l3paper} using the method
described above. Their $\Delta \rho(Q)$ and $D(Q)$ distributions are shown in
Fig.~\ref{fig:l3fig}, both for like-sign and unlike-sign particle pairs.
It can be seen in this figure that an excess at low $Q$ values is observed for
a scenario with inter-$W$ BEC, while no excess is seen in the case where no
inter-$W$ BEC is implemented. The Monte Carlo implementation of the
Bose-Einstein effect was realised by means of the LUBOEI BE$_{32}$ model,
described in~\cite{bec1}, and tuned to the L3 data.

In order to quantify a possible inter-$W$ BEC effect in the data, two methods
are applied. The first consists of integrating the  $\Delta \rho(Q)$
distribution to obtain
\begin{equation}
J=\int_{0}^{Q_{max}} \Delta \rho (Q) dQ,
\end{equation}
\noindent
where the value $Q_{max}$ is taken where the two MC scenarios have converged
to less than one standard deviation.
The obtained value of $J$ for like-sign pairs amounts to
\begin{equation}
J(\pm,\pm)=0.3 \pm 0.33 (stat) \pm 0.15 (syst).
\end{equation}  

In addition, the distribution $D'(Q)$ is computed, defined as
\begin{equation}
D'(Q)=\frac{D(Q)_{\rm data}}{D(Q)_{\rm MC, no inter}},
\end{equation}
\noindent
where the denominator is the $D(Q)$ distribution taken for a MC sample
without inter-$W$ BEC.
This double ratio is used in order to eliminate possible distortions,
introduced by event selections and track mixing.
Finally the $D'(Q)$ distribution is fitted with the following expression
\begin{equation}
D'(Q)=(1+\delta Q)(1+\Lambda \exp(-k^2Q^2)),
\end{equation}
where a non-zero value of $\Lambda$ would indicate the presence of inter-$W$
BEC.
The obtained value of this parameter amounts to
\begin{equation}
\Lambda = 0.008 \pm 0.018 (stat) \pm 0.012 (syst), 
\end{equation}
\noindent
thus being compatible with the absence of an inter-$W$ BEC effect.
\section{DELPHI results}
\begin{figure}
\begin{center}
  \begin{tabular}{cc}
      \mbox{\epsfig{figure=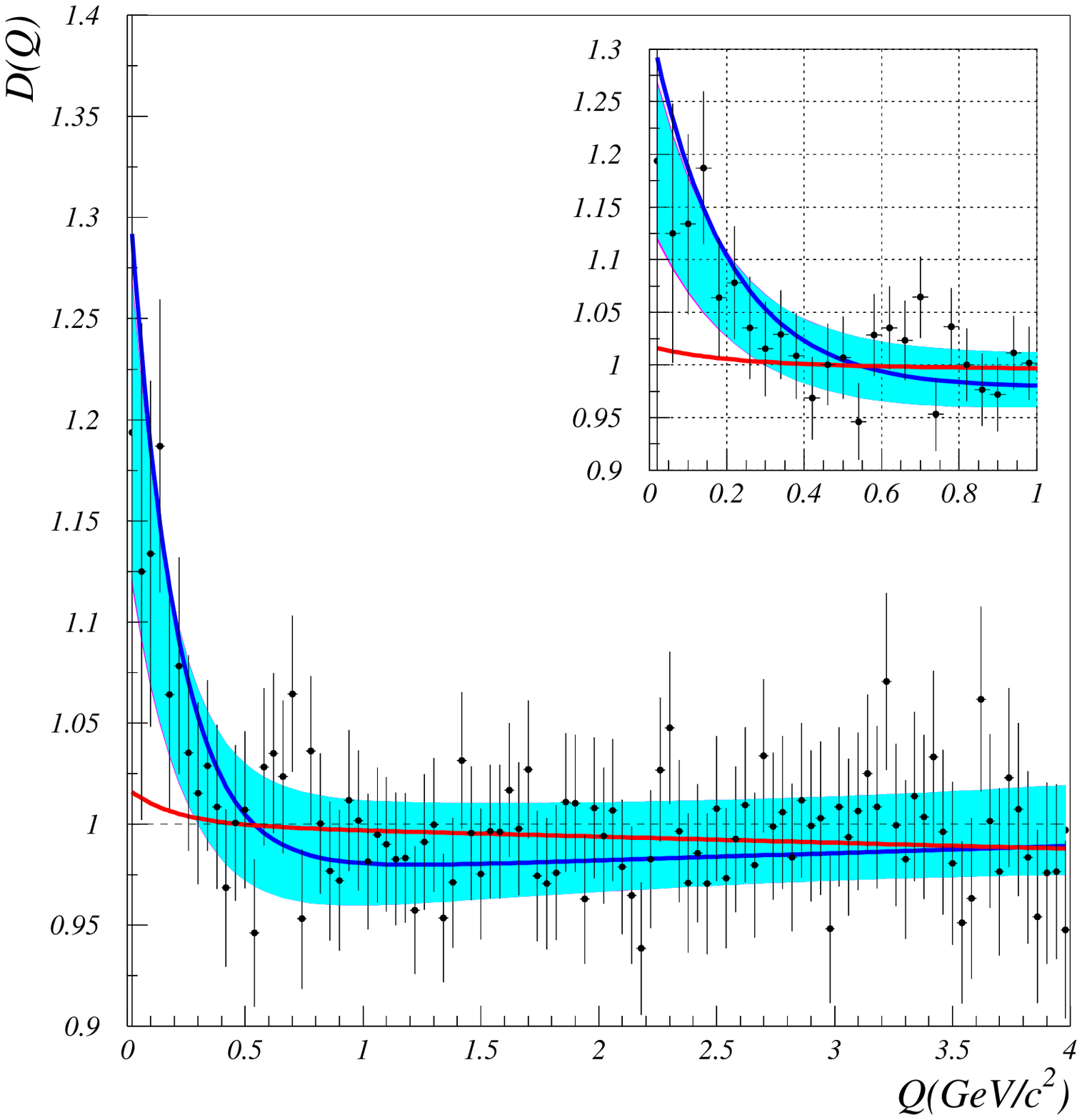,width=7.0cm}}
&
       \mbox{\epsfig{figure=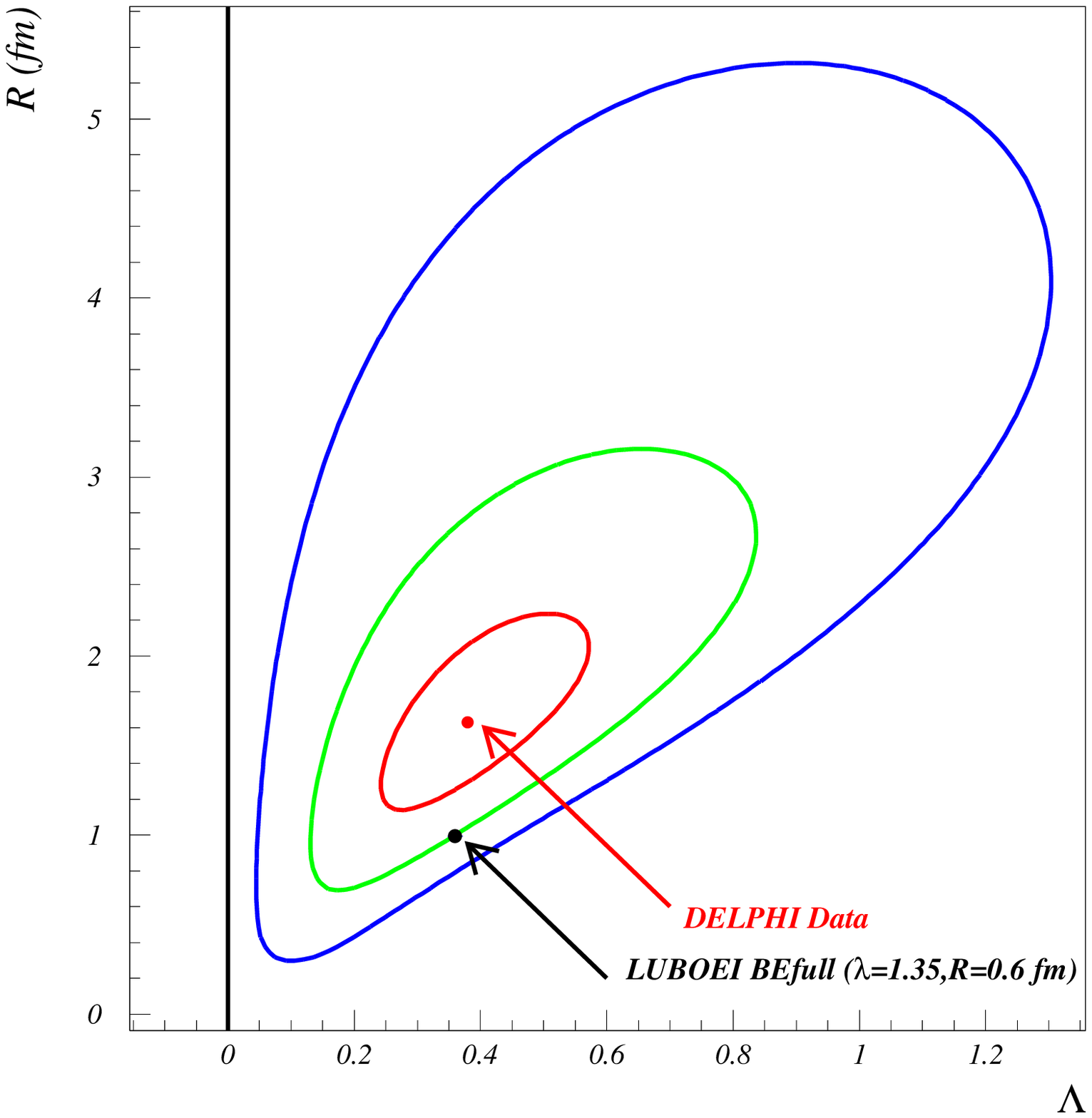,width=7.0cm}}
  \end{tabular}
\end{center}
\caption{The $D(Q)$ distribution for like-sign particle pairs obtained by the
  DELPHI experiment (left). The red line indicates the prediction of a Monte
  Carlo Model including only BEC within one $W$ (BEins). The blue line gives
  the prediction where also BEC between particles from different $W$'s are
  included. The shaded band indicates the result of the fit, using
  Eq.~\ref{dfit} to the real data. On the right, the one, two and three sigma
  contours are given for the fitted $\Lambda$ and $R$ parameters.
\label{fig:delphifig}}
\end{figure}
The DELPHI collaboration has revised their analysis, using the same
formalism as described above. The analysis described in~\cite{dnote} was optimised as a function of the
purity of the selected fully-hadronic $WW$ decays. By increasing the purity of
the selected events, the systematic uncertainty due to the subtraction of
$q\bar{q}(\gamma)$ events decreases, but the amount of statistics is reduced
as well.
The sensitivity of the analysis was further increased by weighing particle pairs,
according to their probability to originate from different $W$'s. This weighing
procedure is model independent and weights can be extracted from data itself,
using mixed events.
Most emphasis is given to a fit of the $D(Q)$ distribution with the following
expression:
\begin{equation}
D(Q)=N(1+\Lambda e^{-R Q})(1+\delta  Q),
\label{dfit}
\end{equation}
\noindent
where the $R$ parameter is either fixed to a value obtained from a MC sample,
using the BE$_{32}$ model, tuned to the DELPHI data, including inter-$W$ BEC,
or where all fit parameters are left free.
The result of both fits can be seen in Fig.~\ref{fig:delphifig}. On the
left-hand side of Fig.~\ref{fig:delphifig} the $D(Q)$ distribution for
like-sign pairs obtained from the data is shown. The blue band indicates the
result of the fit, using Eq.~\ref{dfit} with a fixed $R$ parameter, taking into account
the statistical uncertainties and the correlations between all remaining free
parameters.
The $\Lambda$ value obtained by this approach amounts to
\begin{align}
\Lambda(\pm,\pm) &= 0.241 \pm 0.075 (stat) \pm 0.038 (syst),\\
\Lambda(+,-) &= 0.123 \pm 0.050 (stat) \pm 0.042 (syst),
\end{align}
\noindent
for like-sign and unlike-sign particle pairs respectively.

The BE$_{32}$ model, including BEC between all particles yielded the
following result:
\begin{align}
\Lambda(\pm,\pm) &= 0.360 \pm 0.012 (stat),\\
\Lambda(+,-) &= 0.0785 \pm 0.0057 (stat).
\end{align}

On the right-hand side of Fig.~\ref{fig:delphifig}, the one, two and three
sigma contours of the fitted $\Lambda$ and $R$ parameters are shown. The
central value obtained from the BE$_{32}$ model with inter-$W$ BEC is shown as
well. For like-sign and unlike-sign particle pairs the obtained $\Lambda$
values using this fit are respectively
\begin{align}
\Lambda(\pm,\pm) &= 0.38 \pm 0.16 (stat) \pm 0.038 (syst),\\
\Lambda(+,-) &= 0.131 \pm 0.059 (stat) \pm 0.045 (syst).
\end{align}
\section*{Conclusion}
Both L3 and DELPHI have finalised their analysis, looking for Bose-Einstein
correlations between identical bosons coming from different $W$'s. L3 sees no
indication of the effect, while DELPHI observes an indication of inter-$W$ BEC
for like-sign particle pairs at the level of three standard deviations.
The result of the DELPHI fit indicates that the effect tends to be
concentrated in a very small region in $Q$, indicating that very few pairs
contribute to the effect. This might be good news for the $W$ mass
measurement, where a small amount of particle pairs subjected to the inter-$W$
BEC effect are expected to give a small influence on the reconstructed $W$
mass in the fully hadronic channel.
Results from the OPAL and ALEPH collaborations are eagerly awaited in order
to perform a complete LEP combination of the obtained results.

\end{document}